\documentclass[%
 reprint,
superscriptaddress,
 nofootinbib,
nobibnotes,
amsmath,amssymb, aps,
 showkeys,
]{revtex4-1}

\usepackage{RevTeX_macro_commands}
\usepackage{local_commands}
\usepackage[normalem]{ulem}

\begin{document}
\preprint{APS/123-QED}

\title{Prospects for detection of ultra 
high frequency gravitational waves \intitle{\\}
from hyperbolic encounters with resonant cavities}


\author{Aur\'{e}lien Barrau}
\email{barrau@lpsc.in2p3.fr}
\affiliation{
Laboratoire de Physique Subatomique et de Cosmologie\char`,{} Universit\'e Grenoble-Alpes\char`,{} CNRS/IN2P3\\
53\char`,{} avenue des Martyrs\char`,{} 38026 Grenoble cedex\char`,{} France
}

\author{Juan Garc\'ia-Bellido}
\email{juan.garciabellido@gmail.com}
\affiliation{Instituto de F\'isica Te\'orica UAM/CSIC\char`,{} Universidad Aut\'onoma de Madrid\char`,{} Cantoblanco 28049 Madrid\char`,{} Spain}

\author{Killian Martineau}
\email{martineau@lpsc.in2p3.fr}
\affiliation{
Laboratoire de Physique Subatomique et de Cosmologie\char`,{} Universit\'e Grenoble-Alpes\char`,{} CNRS/IN2P3\\
53\char`,{} avenue des Martyrs\char`,{} 38026 Grenoble cedex\char`,{} France
}

\author{Martin Teuscher}
\email{teuscher@lpsc.in2p3.fr}
\affiliation{
Laboratoire de Physique Subatomique et de Cosmologie\char`,{} Universit\'e Grenoble-Alpes\char`,{} CNRS/IN2P3\\
53\char`,{} avenue des Martyrs\char`,{} 38026 Grenoble cedex\char`,{} France
}
\affiliation{
\'Ecole Normale Supérieure de Paris\char`,{} 45 rue d'Ulm\char`,{} 75005 Paris\char`,{} France
}

\date{April 2024}

\begin{abstract}
In this brief article, we pursue the systematic investigation of possible gravitational wave sources in the gigahertz band. We focus on hyperbolic encounters of light black holes and evaluate precisely the expected signal when accounting for the detailed characteristics of haloscope experiments. Considering the GraHal setup as a benchmark, we insist on the correct signal-to-noise ratio expression, taking into account the appropriate timescales resulting from both physical and instrumental constraints. The associated maximum distance--of the order of a hundredth of an astronomical unit--at which an event can be detected is calculated for optimal, suboptimal, and general trajectories. The main conclusion is that detection seems clearly out of reach.
\end{abstract}

\maketitle


\section{Introduction}
\label{sec:intro}

Observations of gravitational waves (GW) by the LIGO-Virgo collaboration\footnote{Now LIGO-Virgo-Kagra (LVK).} in the [$10$ -- $10^{4}$] Hz range have shaken the science of gravity \cite{KAGRA:2021vkt}. Since then, the field has blown up, arousing a keen interest that does not cease to grow.  At lower frequencies, the LISA project aims at measuring those waves in the [$10^{-4}$--$1$] Hz window \cite{amaroseoane2017laser}. At even lower frequencies, measurements of time delays in pulsar signals by the international pulsar timing array (IPTA) consortium regrouping the North American, European, Indian and Australian  Pulsar Timing Arrays (respectively the NANOGrav, EPTA, InPTA and PPTA collaborations) have recently highlighted lines of evidence for a stochastic gravitational wave background in the nHz range \cite{NANOGrav:2023gor, EPTA:2023fyk, Reardon:2023gzh}. Clearly, most of the research effort is dedicated to the exploration of the low frequency region of the spectrum. 

The high frequency range, however, remains missing in this landscape. There is a good reason for that: no substantial signal is expected from any known astrophysical source, the upper bound being hold by the ringdown signal of neutron star mergers, for which gravitational wave signals of up to 5 kHz are to be expected \cite{Bauswein:2015vxa}. On this specific matter of neutron stars physics, some promising counterarguments were however given in \cite{Casalderrey-Solana:2022rrn}.\\

The absence of known standard sources is obviously not the end of the game as ultra high frequency gravitational wave (UHFGW) signals can be expected from many beyond standard model sources, both in the early or late-time Universe. A review can be found in \cite{Aggarwal:2020olq}. Among those potential sources, light primordial black holes (PBHs) are probably the most natural one~\cite{LISACosmologyWorkingGroup:2023njw}. The gravitational signals they can generate, far above the kHz, either through their evaporation or by merging processes, have already been discussed in \cite{Anantua:2008am, Dong:2015yjs, Franciolini:2022htd, barrau2023prospects, LISACosmologyWorkingGroup:2023njw}. 
In this work we focus on another possibility still in the context of primordial black holes: high frequency GW bursts arising from hyperbolic encounters~\cite{Garcia-Bellido:2017knh,Garcia-Bellido:2017qal,Morras:2021atg,Garcia-Bellido:2021jlq} in dense clusters of PBHs~\cite{Trashorras:2020mwn}. Although open trajectories of this kind could be prevalent, they are much less discussed in the literature. We shall show in the following that there are no good reasons for this when both the speed of the frequency drift and the narrow detector bandwidth are taken into account.

When it comes to detection at frequencies above the kHz, various methods have been considered, spanning from optically levitated dielectric sensors operating in the $[1 , 300]$ kHz range \cite{Arvanitaki:2012cn,Aggarwal:2020umq} to high-energy pulsed lasers at optical frequencies and beyond, up to $10^{19}$ Hz \cite{Vacalis:2023gdz}. These are only examples, and the interested reader can find an exhaustive review in \cite{Aggarwal:2020olq}.
Among those proposals, resonant cavities located at the core of haloscope experiments (originally designed to search for axionic dark matter), can be reused as competitive GW detectors in the GHz range; see \cite{Berlin:2021txa} for an introduction and \cite{barrau2023prospects} for clarifications regarding the importance of the signal time duration. A novel cavity design specifically dedicated to the search for GWs has also recently been proposed \cite{Navarro:2023eii}.\\

The goal of this study is to precisely estimate the distance at which a hyperbolic encounter between light black holes can be observed by a resonant cavity operating in the GHz band. We consider the Grenoble Axion Haloscope (GrAHal) experiment, described in the third section of the article, as a benchmark. It is currently in the commissioning phase and it should be operational in 2024 \cite{9714155, 10359164} \\


In a nutshell, we find that distances that could be reached are way too small to allow any detection. This is important as an intense experimental activity is right now taking place around such ideas.

Nonetheless, we also find that--maybe surprisingly--the sensitivity is not worse (and even better) than for bound trajectories. The fundamental reason behind is that, even for inspirals, the time spent within the experimental bandwidth is very small; hence, the hyperbolic case does not drastically differ from the latter. We will carefully consider different kinds of trajectories (from parabolic to highly eccentric) and different mass ratios. 

\section{General considerations on hyperbolic encounters}

Although highly eccentric trajectories are still to be investigated in details, the case of closed orbits has been intensively considered and the GHz gravitational wave emission is now quite exhaustively understood (see \cite{Franciolini:2022htd, barrau2023prospects}). In principle, hyperbolic encounters should be even more generic--in the sense that unless energy is dissipated it is not possible to gravitationally form a bound system from an unbound one \cite{Garcia-Bellido:2017qal,Garcia-Bellido:2017knh,Jaraba:2021ces,Garcia-Bellido:2021jlq,Morras:2021atg,caldarola2023effects}. In a recent and more abstract study \cite{Teuscher:2024xft}, it was shown that determining the parameters that maximize the strain is not trivial. The most optimal trajectory corresponds to the case where the involved masses are identical, reach the highest possible value allowed by physical constraints and follow a parabola (eccentricity $e=1$). However, if masses are taken at a given lower value, the most favorable trajectory becomes the one with the {\it largest} possible eccentricity.\\

To address the capability of resonant cavities to detect hyperbolic events, the frequency, the strain and the time duration of the phenomenon must be known.\\

To keep the following analysis as general as possible, we do not specify the individual masses $m_1$ and $m_2$ of the black holes and express the results as a function of the reduced mass $\mu = m_1 m_2 / (m_1 + m_2)$ and the total mass $M=m_1+m_2$. Note that $\mu$, and therefore the strain, is maximal when $m_1 = m_2$ (or $\mu = \flatfrac{M}{4}$).\\

The characteristic peak frequency of the burst at periapsis can be written in terms of the periapsis radius $r_p$, which corresponds to the distance to focus at closest approach, and of the gravitational radius $R_S = 2 GM /c^2$  \citep{Teuscher:2024xft}:

\begin{eqnarray}
    f_{\rm p} &=& \frac{1}{2 \pi} \sqrt{\frac{G M (e+1)}{r_p^{3}}} \\ \nonumber
&\approx& 
\SI{1.6}{\giga\hertz} \times \left (\frac{10^{-5} \solarmass}{M} \right ) \left(\frac{R_S}{r_p} \right)^{3/2} \sqrt{\frac{e+1}{2}}~, 
\end{eqnarray}

$c$ being the speed of light and $G$ the gravitational Newton's constant.
We shall argue later, in details, why the periapsis region, that corresponds to the maximum strain emission, is the only relevant zone of the path. 

Interestingly, the previous expression depends only on the total mass of the system $M$, the ratio $R_S/r_p$, and the eccentricity $e$.\\

The typical amplitude for the strain is given by \cite{Garcia-Bellido:2017knh,Teuscher:2024xft} 
\begin{widetext}
\begin{align}\label{eq:strain}
    \mathfrak{h} &\equiv \sqrt{\pref{2}h_{ij}^{\TT}h^{\TT\, ij}} = \sqrt{h_+^2 + h_\times^2} =\frac{\sqrt{2}\,G \mu (GM\w_p)^{2/3}}{D c^4 (e+1)^{4/3}}\left[8 + 13 e^2 + 2 e^4 + 2 e (12 + 5 e^2) \cos{\ihp} + 13 e^2 \cos{2\ihp} + 2 e^3 \cos{3\ihp}\right]^{1/2} \,,
\end{align}
where $D$ is the distance to the source and $\ihp$ the angle along the trajectory. When evaluated at the periapsis, the strain amplitude and the power of the gravitational burst are

    \begin{eqnarray}
    \label{eq:def-hp}
        \mathfrak{h}\subsc{max} &=& \mathfrak{h}_{p} = \frac{2\,\cal{G} \mu (\kappa\w_p)^{2/3}}{Dc^4}\frac{e+2}{(e+1)^{1/3}} \approx 3.6 \times 10^{-25} \times \frac{4 q}{\left( 1+q \right)^2} \frac{G(e)}{G(1)} \left( \frac{M}{10^{-5} \solarmass} \right)^{5/3} \left( \frac{f_p}{\SI{1.6}{\giga\hertz}} \right)^{2/3} \left( \frac{\SI{1}{\mega\rm{pc}}}{D} \right)~, \\
        P\madmax &=& P_p = \frac{32\G\mu^2 \kappa^{4/3} \w_p^{10/3}}{5c^5(e+1)^{2/3}}\approx 2.4 \times 10^{24} L_\odot \times \frac{1}{(e+1)^{2/3}} \left( \frac{4 q}{\left( 1+q \right)^2} \right)^2\left( \frac{M}{10^{-5} \solarmass} \right)^{10/3} \left( \frac{f_p}{\SI{1.6}{\giga\hertz}} \right)^{10/3}~,
    \end{eqnarray}
    \end{widetext}

where $L_\odot=\SI{3.828d26}{\watt}$ is the solar luminosity. In the above the function $G(e) \equiv \flatfrac{(e+2)}{(e+1)^{1/3}}$, parameter $\kappa=\cal{G}M$ and ratio $q \equiv m_1 / m_2$ have been introduced.\\




    From now on, the pulsation $\omega_p$ of the generated burst at the periapsis is set equal to the detector frequency $\nu$, such that $\omega_p = 2 \pi \nu$. \\

The time duration of the signal in a detector bandwidth $\Delta \nu$ centered around an operating frequency $\nu$ can be computed using the conservation of momentum \cite{Garcia-Bellido:2017knh, Teuscher:2024xft}
\begin{equation}
   t_{\Delta \nu}(\nu, \Delta \nu, e) = \frac{1}{\pi \nu} \sqrt{1+\frac{1}{e}} \sqrt{\frac{\Delta \nu}{\nu}} ~.
   \label{eq:tdeltanu}
\end{equation}

Since $\nu$ and $\Delta \nu$ are fixed by the experimental apparatus, this time duration depends only on the eccentricity $e\geq 1$ of the hyperbolic trajectory. Notice however that $1 < \sqrt{1+e^{-1}} \leqslant \sqrt{2} $, so this dependence is weak.

\section{Experimental setup}

One of the most promising possibilities to search for GWs around the GHz is to reuse resonant cavities located inside haloscope instruments, originally dedicated to axion-like particle searches \cite{Berlin:2021txa, Aggarwal:2020olq}. For signals that would remain coherent for very long times, already existing experiments exhibit impressive sensitivities on the strain, reaching $h \sim 10^{-22}$ \cite{Berlin:2021txa}. It was however emphasized in \cite{Franciolini:2022htd, barrau2023prospects} that the time $t_{\Delta\nu}$ spent by the signal within the sensitivity band of the detector is crucial when estimating the signal to noise ratio (SNR). In principle, maximizing the strain is not necessarily equivalent to maximizing the SNR. For haloscope experiments, the usual Dicke radiometer formula \cite{Sikivie:2020zpn, Berlin:2021txa, Berlin:2022hfx} reads 
\begin{equation}
    \SNR \sim \frac{P\subsc{sig}}{k_B T\subsc{sys}}\sqrt{\frac{\teff}{\Delta\nu}} ~,
    \label{eq:SNRdef}
\end{equation}
$P\subsc{sig}$ being the signal power, $k_B$ the Boltzmann constant, $T\subsc{sys}$ the temperature of the system (that includes all contributions), $\Delta \nu$ the spectral bandwidth, and $\teff$ the effective time during which the detector gathers enough signal. The expression of this effective time is discussed in detail below.

The power associated with GW signals has been estimated to be \cite{Berlin:2021txa}
\begin{equation}
\label{eq:def-Psig}
    P\subsc{sig} = \pref{2\mu_0 c^2} V\subsc{cav}^{5/3}(\eta B_0)^2 \Qeff (2\pi\nu)^3 \mathfrak{h}^2~,
\end{equation}
in which $\mu_0$ is the vacuum magnetic permeability and $V\subsc{cav}$, $\eta$, $B_0$, $\Qeff$ are respectively the volume of the cavity, the coupling coefficient between gravitational waves and resonance modes, the static magnetic field inside the cavity, and an effective quality factor. Usually $\Qeff$ is taken to be the \textit{cavity quality factor} $Q=\flatfrac{\nu}{\Delta\nu}$. Care must however be taken when the time $t_{\Delta\nu}$ spent by the signal in the bandwidth is smaller than $\tmin \equiv \flatfrac{1}{\Delta\nu}$. In this regime $\Qeff$ is no longer the cavity quality factor $Q$ but the \textit{signal quality factor} \cite{Kim:2020kfo, barrau2023prospects}
\begin{equation}
    \label{eq:-Q-and-Qid}
    \Qeff = \frac{t_{\Delta\nu}}{\tmin} Q = \nu \times t_{\Delta\nu} < Q.
\end{equation}

Equations \eqref{eq:SNRdef} and \eqref{eq:def-Psig} raise an important point: \textit{the quantity to maximize, at a fixed given frequency, in order to reach the best sensitivity is} $\mathfrak{h}^2\sqrt{\teff}$ \textit{rather than $\mathfrak{h}$ itself.}  We have however shown in \cite{Teuscher:2024xft} that, in practice, for hyperbolic trajectories, the highest signal (in the sense of $\mathfrak{h}^2\sqrt{\teff}$) is always reached at the periapsis.\\


Numerical estimates of the observational distance require the experimental quantities of interest to be specified. To do so, we use the GrAHal experiment as a benchmark. Initially designed for the search
of axionic dark matter in our galactic halo, GrAHal is a haloscope platform located in Grenoble \cite{Grenet:2021vbb, GrAHall-CAPP-2024} whose development has been motivated by the existence on site of a 43 Tesla multiconfiguration hybrid magnet. 
Table \ref{Table GrAHal configurations} presents the parameters relevant for this study, for five different configurations accessible at the GrAHal platform. Introducing $T_{\text{cav}}$ as the cavity temperature, the benchmark for the system temperature appearing in this table corresponds to the sum of three terms: the amplifier noise temperature, which dominates, the cavity thermal noise $h \nu /\left(\text{exp} \left[h \nu / k_B T_{\text{cav}}\right]-1 \right)$, and the zero-point fluctuations of the blackbody gas. The frequency of the transverse-magnetic mode $\nu_{TM010}$ provided in this table will, from now on, be considered as the operating frequency of the detector $\nu = \flatfrac{\w_p}{2\pi} = \nu_{TM010}$. 


\begin{table}[h!]
\begin{center}
\begin{tabular}{|c|c|c|c|  }  
 \hline
 \multicolumn{4}{|c|}{GrAHal configurations} \\
 \hline
$B_{0} (\text{T})$ & Cavity volume ($\text{m}^3)$ & $T_{\text{sys}}$ $(\text{K})$ & $\nu_{TM010} (\text{GHz})$ \\
 \hline
9   & $5.01\times10^{-1}$    & 0.3 &   0.34 \\
 17.5 &   $3.22\times10^{-2}$   & 0.3 & 0.79 \\
 27 & $1.83\times10^{-3}$  & 0.4 &  2.67 \\
 40    & $1.42 \times10^{-4}$  & 1.0 &  6.74 \\
 43 &   $4.93 \times10^{-5}$   & 1.0 & 11.47 \\
 \hline
\end{tabular}
\caption{Main possible configurations for GrAHal using the 43 T Grenoble hybrid magnet \cite{9714155, 10359164}.
}
\label{Table GrAHal configurations}
\end{center}
\end{table}

To perform sensitivity estimates, the effective time $\teff$ appearing in the signal-to-noise ratio has to be specified. It depends on the time $t_{\Delta \nu}$ during which the evolving gravitational wave frequency drifts into the sensitivity bandwidth of the experiment $\Delta \nu$, but also on timescales relevant for the experiment, such as the maximal integration time $t_{\text{max}}$ or the inverse sampling rate $t_{\text{min}} = \flatfrac{1}{\Delta\nu}$.\\ 

We have shown in \cite{barrau2023prospects} that three regimes should \textit{a priori} be distinguished depending on the hierarchy between $t_{\text{min}}$, $t_{\text{max}}$, and $t_{\Delta \nu}$. It is however easy to show that the condition  $t_{\Delta \nu} < t_{\text{min}}$ is equivalent to $Q^{3/2}\gtrsim 1$ for hyperbolic trajectories, which is obviously satisfied. The case in which the signal duration in the detector bandwidth is smaller than the time separating two successive samples is therefore the only regime which needs to be considered in this work and the corresponding loss in SNR is taken into account by setting \cite{barrau2023prospects}
\begin{equation}
\label{eq:teff-def}
    \teff = \flatfrac{t_{\Delta\nu}^2}{t_{\text{min}}}\mperiod
\end{equation}

Numerical estimates are obtained using Eq. (\ref{eq:tdeltanu}) in the specific case of a resonant cavity. This leads to
\begin{equation}
   t_{\Delta \nu} \approx \SI{5e-13}{\second}  \fiducial{\SI{2.67}{\giga\hertz}}{\nu}{}{} \fiducial{10^5}{Q}{}{1/2}\left(\frac{1+e^{-1}}{2}\right)^{1/2} ~.
   \label{time duration signal}
\end{equation}

\section{Maximal distance accessible}

The estimation of the maximum accessible distance, basically obtained by requiring SNR $>1$, depends both on the sensitivity of the experiment on the strain and duration of the emitted signal. \\

\subsection{General considerations and strain sensitivity}

Before performing the detailed estimates, it is worth underlying an important point to emphasize more clearly why we focus on the periapsis throughout all this study.\\

In \cite{barrau2023prospects}, it has been shown that, in the case of a coalescing system, the sensitivity on the strain is maximal for the {\it lowest} black hole masses and decreases as the chirp mass of the system increases. This simply reflects the fact that the lower the mass, the earlier the binary system is considered in its inspiral phase, and the longer the signal remains inside the detector bandwidth. Focusing on the dynamics long before the merging, it is always possible to find parameters such that the signal spends an arbitrary large amount of time within the experiment bandwidth. In this case, the required strain to get a measurable signal obviously gets smaller.

Of course, this does not mean that lower masses are more likely to be observed as the emitted signal decreases when the chirp mass decreases. The increase in $t_{\Delta \nu}$ never comes close to compensate for the decrease in strain (see Figs.~4 and 5 of \cite{barrau2023prospects}).\\


In the case of hyperbolic encounters, the same argument could, in principle, be made. The instantaneous (mean) value of the frequency of the signal  
does indeed decrease continuously when the system is considered far away from the periapsis. In this case, speeds are smaller, hence the system stays longer in the bandwidth. The frequency at the periapsis should then be much above the experiment frequency. However, in addition to the already drastic decrease in SNR described in \cite{barrau2023prospects} and mainly due to the necessarily smaller masses involved in the process, two additional effects make this situation even more marginal for open trajectories than for bounded ones. First, as soon as the trajectory is considered far away from the periapsis, the Fourier transform of the signal gets wider and the amount of signal within the bandwidth of the experiment decreases. Second, the trajectory becomes nearly linear and, therefore, even at fixed masses, the signal would be strongly damped.\\

For all those reasons, we consider only the signal emitted around the periapsis. As the system basically stays here for half a period, it is impossible that it spends a vast amount of time within the bandwidth while emitting UHFGWs.\\


Plugging Eqs. (\ref{eq:tdeltanu}) and (\ref{eq:def-Psig}) into Eq. (\ref{eq:SNRdef}) it is straightforward to estimate the minimum strain that can be detected: 

\begin{widetext}
\begin{eqnarray}
   \mathfrak{h}_{\text{min}} &=& \sqrt{\frac{\mu_0 c^2 k_B}{4\pi}}\text{SNR}^{1/2} T_{\text{sys}}^{1/2}  V_{\text{cav}}^{-5/6} \eta^{-1} B_0^{-1} \nu^{-1} Q^{1/2} \\\ \nonumber
   &\simeq& 1.6\times 10^{-12}\ \fiducial{\text{SNR}}{1}{}{1/2}
   \fiducial{T_{\text{sys}}}{0.4}{K}{1/2}
   \fiducial{1.83 \times 10^{-3}~\text{m}^3}{V_{\text{cav}}}{}{5/6} 
   \fiducial{0.1}{\eta}{}{}
   \fiducial{27 ~\text{T}}{B_0}{}{} 
   \fiducial{2.67~\text{GHz}}{\nu}{}{}
   \fiducial{Q}{10^5}{}{1/2}~.
\end{eqnarray}
\end{widetext}

It is important to notice that, for standard fiducial parameters, the numerical value obtained here is far above the sensitivity around $10^{-22}$ for long time coherent signals \cite{Berlin:2021txa}. The reason for this apparent loss in sensitivity is obviously not the experiment itself, but the extremely small time duration of the signal, as reflected by Eq. (\ref{time duration signal}). As discussed previously, considering parts of the hyperbolic trajectory which increase $t_{\Delta \nu}$ would decrease the required strain for detection but would still dramatically worsen the distance at which an event could be seen. 
The value of $h_{\text{min}}$ calculated in this work is consistent with the results we have previously obtained for coalescing systems \cite{barrau2023prospects}. Should the orbit be elliptic or hyperbolic, the time spent by the signal in the (very high frequency) detector bandwidth would be roughly the same (as it is anyway smaller than the orbital period) as soon as one focuses on the most favorable case, that is the largest masses.\\  

\subsection{Description of the parameter space}

\begin{figure}[!h]
    \centering
    \includegraphics[width=0.48\textwidth]{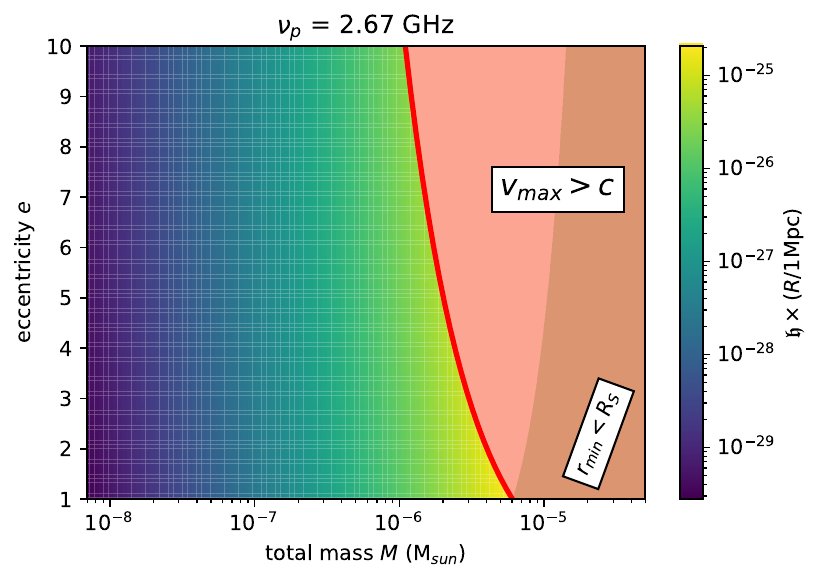}
    \caption{Emitted strain $\mathfrak{h}$ as a function of eccentricity and total mass of the system, for a frequency at periapsis $\omega_p/(2\pi)$ fixed at the detector frequency $\nu = 2.67$ GHz and considering $m_1 = m_2$. The solid red line $e = e_*(M,\omega_p)$ is the dividing line between physically allowed and excluded regions of the parameter space, considering two constrains: the "no-merging" constrain $r_{\text{min}} > R_S$ and the ``maximal  allowed speed" constrain $v_{\text{max}} < c$. As demonstrated in section III B of \cite{Teuscher:2024xft}, the intersection between this dividing line and the $e = 1$ line is the point of the parameter space where $\mathfrak{h}$ is maximal. When one variable is fixed, the strain evolves as $\mathfrak{h} \propto M^{5/3}$ or $\mathfrak{h} \propto e^{2/3}$ (for sufficiently large values of $e$).}
    \label{fig:strainvsMande}
\end{figure}

The typical strain $\mathfrak{h}$, normalized at a distance of 1 Mpc, is plotted as a function of eccentricity and total mass in Fig. \ref{fig:strainvsMande}, assuming $m_1 = m_2$ and a periapsis frequencies given by typical values of the $\nu_{TM010}$ mode of GrAHal. As this work is specifically focused on signals in resonant cavities and not on the mathematical aspects of  GW emission along the trajectory, we refer the interested reader to \cite{Teuscher:2024xft} to get details about the ingredients necessary for this plot. \\

We impose two different constrains on the system:

\begin{itemize}
    \item the relative velocity between the two masses should not exceed the speed of light, i.e., $v_{\text{max}}<c$;
    \item the system should not merge, i.e., $r_{\text{min}} > R_S$.
\end{itemize}

Under those constrains, all eccentricities are bounded by
\begin{equation}
\label{eq:bound-on-e}
    e \leqslant e_* =\frac{c^3}{G M \w_p} - 1~.
\end{equation}

The accessible masses are therefore given by
\begin{equation}
    M \leq 6.1 \times 10^{-6} \solarmass \left( \frac{2}{e+1} \right) \left(\frac{2\pi\times2.67 ~\text{GHz}}{\omega_p} \right) ~,
\end{equation}

the highest value corresponding to the parabolic trajectory $e=1$.\\

We now consider three different relevant situations:

\begin{itemize}
    \item The so-called ``general case", filling almost all the allowed region of the parameter space;
    \item The ``optimized trajectories", that correspond to trajectories that maximize the strain at fixed $e$ or $M$. These are the trajectories that lie along the curve corresponding to the constraint $e = e_*$;
    \item The ``parabolic trajectory" given by $e_*=1$. Even though being a single point in the previous parameter space, we still discuss it as it is trajectory that  saturates the constraint $e = e_*$ while corresponding to the highest $M$ accessible. As a result, this trajectory is the one generating the highest strain.
\end{itemize}

\subsection{General case}


Plugging the previous formulas Eqs. (\ref{eq:def-hp}) and (\ref{eq:tdeltanu}) for the strain and time duration of the signal into the expression of the SNR (\ref{eq:SNRdef}), one is led to a maximal distance for a detectable event given by
\begin{eqnarray}
    D_{\text{max}} &=& \frac{2^{2/3}\pi^{13/6}}{\sqrt{\SNR}}\frac{V\subsc{cav}^{5/6}\eta B_0 }{\sqrt{\mu_0 k_B T\subsc{sys}}} \frac{4\mu}{M} \left(\frac{\cal{G}M\nu}{c^3}\right)^{5/3}  \\ 
    \nonumber
    &&\times\ Q^{1/4}\Qeff^{1/2} \left(\teff\nu\right)^{1/4}\frac{e+2}{(e+1)^{1/3}}.
\end{eqnarray}


This is the most general and concise equation one can get. Replacing $\Qeff$ and $\teff$ using Eqs. \eqref{eq:-Q-and-Qid} and \eqref{eq:teff-def}, it can be written more conveniently as
\begin{eqnarray}
\label{eq:generalfiducial}
    D_{\text{max}} &=& \frac{2^{2/3}\pi^{7/6}}{\sqrt{\SNR}}\frac{G^{5/3}}{c^5 \sqrt{\mu_0 k_B}}\frac{V\subsc{cav}^{5/6}\eta B_0 \nu^{5/3} Q^{-1/2}}{T\subsc{sys}^{1/2}} \\ 
    \nonumber
    && \times \frac{4\mu}{M} M^{5/3} H(e)\\ 
    \nonumber
    &=& \frac{\SI{2.4e3}{\kilo\meter}}{\sqrt{\SNR}}
  \fiducial{V\subsc{cav}}{1.83d-3}{\meter\cubed}{5/6}
  \fiducial{\eta}{0.1}{}{}
    \fiducial{B_0}{27}{\tesla}{} \times \\
  \nonumber
 &&  \fiducial{\nu}{2.67}{\giga\hertz}{5/3}
    \fiducial{Q}{10^5}{}{-1/2}
    \fiducial{T\subsc{sys}}{0.4}{\kelvin}{-1/2} \times  \\
   && \left( \frac{H(e)}{H(1)} \right)
    \fiducial{M}{6.1d-8}{\solarmass}{5/3} \left(\frac{4\mu}{M}\right),   
\end{eqnarray}
where "AU" stands for astronomical unit and 
\begin{equation}
    H(e) \equiv \dps \frac{e+2}{(e+1)^{1/3}}\left(1+\frac{1}{e}\right)^{1/2}~.
\end{equation}
For all $e \geq 1$, $H(e)$ is a monotonously increasing function of $e$ that takes values $H(1)\simeq 3.4$ and $H(10)\simeq 5.7$.
The $4 \mu / M$ factor becomes one when the two black hole masses have the same mass (this situation also corresponds to the one that maximizes the emitted strain $\mathfrak{h}$).

We have here taken into account two important effects required to use the Dicke formula in this extreme regime: the reduction of the effective time due to the limited sampling rate and the modification of the quality factor implied by the partial loading of the cavity.


\subsection{Optimized trajectories}

In this section, we focus on favorable situations.\\ 

As demonstrated in \cite{Teuscher:2024xft}, and somehow counter intuitively, at fixed total mass $M$, the trajectory that maximizes the SNR keeping the constraint that the periapsis frequency is fixed at the detector frequency (as in the III. B. case of \cite{Teuscher:2024xft}), is the one with the \textit{highest} eccentricity.\\

Those trajectories are such that the black hole masses and the periapsis frequency are not independent variables: they must saturate Eq. \eqref{eq:bound-on-e}. If the gravitational wave frequency at the periapsis is taken to be the one of the detector, this fixes the total mass of the observed system such that
\begin{eqnarray}
\label{eq:relation-M-freq}
    M &=& \frac{c^3}{(e+1) 2 \pi G\nu_p}\\
    &=& \SI{1.2d-5}{\solarmass} \left(\frac{1}{e+1} \right) \left(\frac{\SI{2.67}{\giga\hertz}}{\nu}\right).
\end{eqnarray}


The maximal distance at which en event is accessible becomes
\begin{eqnarray}
    D_{\text{max}} &=& \frac{1}{2\sqrt{\pi}\sqrt{\SNR}}\frac{V\subsc{cav}^{5/6}\eta B_0}{\sqrt{\mu_0 k_B T\subsc{sys}}}\frac{4\mu}{M} Q^{-1/2} J(e)\\ 
    \nonumber
    &=& \frac{\SI{5.1e6}{\kilo\meter}}{\sqrt{\SNR}}\left(\frac{4\mu}{M}\right)
  \fiducial{V\subsc{cav}}{1.83d-3}{\meter\cubed}{5/6} \fiducial{\eta}{0.1}{}{}\times \\
 && \fiducial{B_0}{27}{\tesla}{}
    \fiducial{T\subsc{sys}}{0.4}{\kelvin}{-1/2}
    \fiducial{Q}{10^5}{}{-1/2} \left( \frac{J(e)}{J(1)}\right),
\end{eqnarray}

with $J(e) \equiv \dps \frac{e+2}{e^{1/2}(e+1)^{3/2}}$ and $J(e) \leqslant J(1)$.\\


Let us now focus on the ``most optimal" trajectory, where in addition to satisfying Eq. \eqref{eq:relation-M-freq} (which corresponds to the case where the involved masses--assumed equal--reach the highest possible value allowed by physical constraints), the trajectory forms a parabola (eccentricity $e=1$). Inserting these constraints in Eq. \eqref{eq:generalfiducial} yields

\begin{eqnarray}
\label{eq:finalvalue}
    D_{\text{max}} &=& \frac{3}{2^{5/2} \pi^{1/2}\sqrt{\SNR}}\frac{V\subsc{cav}^{5/6}\eta B_0}{\sqrt{\mu_0 k_B T\subsc{sys}}} \frac{4\mu}{M} Q^{-1/2}
 \\
   &=& \frac{\SI{3.4e-2}{\rm{A.U.}}}{\sqrt{\SNR}}\left(\frac{4\mu}{M}\right)
  \fiducial{V\subsc{cav}}{1.83e-3}{\meter\cubed}{5/6} \\ \nonumber
 && \times \fiducial{\eta}{0.1}{}{}
    \fiducial{B_0}{27}{\tesla}{}
    \fiducial{T\subsc{sys}}{0.4}{\kelvin}{-1/2}
    \fiducial{Q}{10^5}{}{-1/2}.
\end{eqnarray}\\

\subsection{Discussion}

The previous expressions clearly exhibit the importance of each effect on the final reachable distance, which is very substantially decreased when compared to the first ideal case. This is especially interesting as, in some studies, even this ``ideal" case is not treated with care and the physical duration of the signal itself is over-estimated.\\


For fiducial parameters close to those expected for the GrAHal experiment, the maximum distance at which an event could be detected is of the order of a few hundredths of astronomical unit in the best case. This is a tiny result. 



\section{Conclusion}
\label{sec:conclu}

In this brief article, we have investigated the detection capability of resonant cavities operating in the GHz range for hyperbolic encounters of light black holes. It might have been expected that the maximum distance at which an event can be detected is much less than in the case of a bound orbit for which the signal is assumed to be integrated over many cycles. In a hyperbolic encounter, the considered frequency (and amplitude) it attained only during a fraction (typically one half) of the associated orbital period. The naive reasoning is however misleading as, even for a closed orbit, the frequency drift is so fast--for the highest masses accessible, which are the most relevant ones--that it is anyway measured during a timescale smaller than an orbital period. This is why the distances found in this work are substantial when compared to those previously estimated for coalescing systems. Slight differences come from considering black holes grazing each other (in the hyperbolic case) instead of being at the ISCO (in the circular case). A more important difference is the following. When dealing with circular orbits, one assumes that the frequency drift is {\it only} due to the emission of gravitational waves. This is obviously the way to go as the signal would otherwise be monochromatic and stationary: the Newtonian trajectory does not lead to any frequency evolution. The other way around, in this study, we have assumed that the evolution of the frequency is {\it only} due to the evolution of $\dot{\phi}$ along the (classical) trajectory. As explained in \cite{Teuscher:2024xft}, this is also legitimate as the backreaction of the gravitational wave emission is usually small when the considered masses do not saturate the associated bound. In principle, this should however be investigated in more details. In the optimum case that is emphasized here, this makes our point even stronger as this means that the calculated distance is actually over-estimated. This is an upper bound on $D_{\text{max}}$ as the actual instantaneous frequency of the signal leaves the bandwidth of the detector even faster due to backreaction.\\


For typical fiducial parameters, the distance is decreased to a few thousands of kilometers. We therefore conclude that, contrarily to what was initially hoped, using resonant cavities to detect gravitational waves from black holes is not currently realistic. Especially when taking into account that the situation studied here is--at least in the way it is treated--more favorable than the usual inspiralling one.\\

Still, in a future work, beyond the scope of this paper, it could be welcome to estimate the actual shape and drift of the signal when backreaction is accurately accounted for.

\appendix
\section{Separately accounting for different effects on the maximal detection distance}
\label{appx:-hcross}

For pedagogical purposes, we now detail the roles played by the different effects modifying the naive Dicke formula.\\

{\paragraph{Ideal measurement}
$\text{}$}\\


Let us first ignore the subtleties previously raised and assume that the effective time of the signal is only determined by the physical drift in time of the frequency while the effective quality factor remains the cavity one. In this case the maximum reachable distance reads

\begin{eqnarray}
    D_{\text{max}} &=& \frac{3\pi^{1/4}}{2^{23/8} \sqrt{\SNR}}\frac{V\subsc{cav}^{5/6}\eta B_0}{\sqrt{\mu_0 k_B T\subsc{sys}}} \frac{4\mu}{M} Q^{5/8}
 \\
   &=& \frac{\SI{2.6e4}{\rm{A.U.}}}{\sqrt{\SNR}}\left(\frac{4\mu}{M}\right)
  \fiducial{V\subsc{cav}}{1.83e-3}{\meter\cubed}{5/6} \\ \nonumber
 && \times \fiducial{\eta}{0.1}{}{}
    \fiducial{B_0}{27}{\tesla}{}
    \fiducial{T\subsc{sys}}{0.4}{\kelvin}{-1/2}
    \fiducial{Q}{10^5}{}{5/8}.
\end{eqnarray}\\

{\paragraph{Limitation from sampling time}
$\text{}$}\\


To take into account the smallness of the time spent by the signal within the bandwidth compared to the sampling time, the effective time $\teff$ becomes $\flatfrac{t_{\Delta\nu}^2}{\tmin}$ and the expression is changed to

\begin{eqnarray}
    D_{\text{max}} &=& \frac{3}{2^{11/4} \sqrt{\SNR}}\frac{V\subsc{cav}^{5/6}\eta B_0}{\sqrt{\mu_0 k_B T\subsc{sys}}} \frac{4\mu}{M} Q^{1/4}
 \\
   &=& \frac{\SI{2.8e2}{\rm{A.U.}}}{\sqrt{\SNR}}\left(\frac{4\mu}{M}\right)
  \fiducial{V\subsc{cav}}{1.83e-3}{\meter\cubed}{5/6} \\ \nonumber
 && \times \fiducial{\eta}{0.1}{}{}
    \fiducial{B_0}{27}{\tesla}{}
    \fiducial{T\subsc{sys}}{0.4}{\kelvin}{-1/2}
    \fiducial{Q}{10^5}{}{1/4}.
\end{eqnarray}\\

{\paragraph{Limitation from cavity loading}
$\text{}$}\\

If the incomplete cavity loading is now accounted for, the cavity quality factor $Q$ is replaced by the signal quality factor $Q\times \flatfrac{t_{\Delta\nu}}{\tmin}$ and the distance becomes


\begin{eqnarray}
    D_{\text{max}} &=& \frac{3}{2^{21/8} \pi^{1/4}\sqrt{\SNR}}\frac{V\subsc{cav}^{5/6}\eta B_0}{\sqrt{\mu_0 k_B T\subsc{sys}}} \frac{4\mu}{M} Q^{-1/8}
 \\
   &=& \frac{\SI{3.1}{\rm{A.U.}}}{\sqrt{\SNR}}\left(\frac{4\mu}{M}\right)
  \fiducial{V\subsc{cav}}{1.83e-3}{\meter\cubed}{5/6} \\ \nonumber
 && \times \fiducial{\eta}{0.1}{}{}
    \fiducial{B_0}{27}{\tesla}{}
    \fiducial{T\subsc{sys}}{0.4}{\kelvin}{-1/2}
    \fiducial{Q}{10^5}{}{-1/8}.
\end{eqnarray}
\\

{\paragraph{Cumulating all effects}
$\text{}$}\\


The correct description of the phenomenon requires to take into account simultaneously all the previously mentioned effects. This provides the result obtained in Eq. \eqref{eq:finalvalue}.

\bibliography{references}

\end{document}